\documentclass[aps,pra,twocolumn,amsfonts, amssymb, amsmath, amsthm,groupedaddress]{revtex4-1}
\usepackage{url}
\usepackage{color}
\usepackage{graphicx}
\usepackage{amsthm}

\newcommand{\bra}[1]{{\langle#1|}}
\newcommand{\ket}[1]{{|#1\rangle}}
\DeclareMathOperator{\Tr}{\operatorname{Tr}}
\begin{document}
%\preprint{}

\title{Quantum Steganography over Noisy Channels: Achievability and Bounds}


\author{Chris Sutherland$^1$ and Todd A. Brun$^{1,2}$}
%\thanks{}
\affiliation{$^1$Department of Physics,
$^2$Ming Hsieh Department of Electrical Engineering, \\
University of Southern California, Los Angeles, California}


\date{\today}

\begin{abstract}
Characterizing secret communication over noisy quantum channels is an interesting problem from both a practical and theoretical perspective. Suppose Alice and Bob wish to communicate secret information so that an eavesdropper Eve will not suspect any type of encoded communication between the two. Classical or quantum cryptography will not suffice since it is always clear secret communication is taking place. Therefore Alice and Bob must execute what is known as a quantum steganographic protocol. Assuming Eve only has partial knowledge of the channel connecting Alice and Bob, we show that for the bit-flip and depolarizing channels Alice can use Eve's lack of knowledge of the channel parameter to encode quantum information steganographically. We give an explicit encoding procedure and calculate the rate at which Alice and Bob can communicate secretly. We also show that our encoding is optimal for nondegenerate quantum codes. We calculate the rate at which secret key must be consumed. Finally, we discuss the possibility of steganographic communication over more general quantum channels, and conjecture a general formula for the steganographic rate.
\end{abstract}

%\keywords{}

\maketitle

\section{Introduction}\label{intro}
Suppose Alice and Bob are the respective leaders of two countries and they wish to communicate highly classified information with each other over a public channel. They do not want other countries to know they are communcating secret information, perhaps because they have a history of shady international political relations. Simple cryptography would not be good enough here beacause it would alert a potential eavesdropper (Eve) that secret communication is taking place, even if she cannot read it. Therefore, if Alice and Bob wish to keep their conversations secret, they must employ a steganographic protocol.

Both cryptography and steganography are interesting and well-developed subjects, the studies of which date back millenia \cite{greekstego, singh2000code}. In cryptography, a secret message is encrypted using a shared secret key between Alice and Bob, and Alice sends the resulting \textit{ciphertext} across a channel to be decoded. Should Eve observe this ciphertext, she would not be able to decode it without the secret key. However, she would undoubtedly become suspicious if she weren't already, due to Alice sending encrypted messages to Bob.

Steganography solves this problem of secrecy. Although cryptography allows for \textit{secure} communication, in this paper we are interested in \textit{secret} communication. In steganography, a secret message is hidden into a larger \textit{covertext}, which appears to Eve as an innocuous message. This seeming innocuousness of the message is what makes the protocol secret. The hidden message may also be encrypted itself to make the protocol not only secret but secure, so that even if Eve were tipped off to there being secret communication between Alice and Bob, she would not be able to decode the hidden message. For example, digital audio, video, and pictures are increasingly furnished with distinguishing but imperceptible marks, which may contain a hidden copyright notice or encrypted serial number \cite{petitcolas1999information}.

Ever since Shor's remarkable discovery that a quantum computer could solve the prime factorization problem efficiently, hence cracking one of the internet's most common encryption schemes \cite{shor1999polynomial}, interest in quantum cryptography has been intense. Quantum steganography is of more recent development \cite{natoristego,banerjeestego,gea2002hiding}. The protocol we will be considering is to encode quantum information steganographically as error syndromes of a quantum error-correcting code. This was detailed extensively by Shaw and Brun \cite{shaw2010hiding, shaw2011quantum}, where it was shown that such schemes can hide both quantum and classical information, with a quantitative measure of secrecy, even in the presence of a noisy physical channel. 

A more precise analysis of this quantum steganographic protocol over noiseless channels was done by the present authors in terms of achievability and converse proofs \cite{sutherland2018quantum}. In this work, we treated the case where Eve believes the channel connecting Alice and Bob to be some noisy quantum channel, but the actual physical channel is noiseless, and we gave optimal rates of steganographic communication. 

A related field is known as covert quantum communication~\cite{qcovert1, qcovert2, qcovert3, qcovert4, qcovert5}. In covert communications, it is often assumed that the channel between Alice and Bob is a noisy optical channel, which is modelled by a beamsplitter with some transmissivity parameter that characterizes how many photons are lost to Eve. Covert quantum communication can be seen as a special case of quantum steganography over noisy quantum channels in the case where the eavesdropper has exact knowledge of the channel, and where Eve assumes the channel is idle (so only noise is being transmitted). Similarily, quantum steganography is a type of covert quantum communication where Eve knows about the covertext communication but not the hidden stegotext, and where Eve may not have perfect knowledge of the channel. In covert quantum communication, it has been shown that in general one can secretly communicate an amount of classical information which scales like the square root of the number of channel uses.

The goal of this paper is to extend the previous work on quantum steganography over noiseless channels in \cite{sutherland2018quantum} to the scenario where the channel Alice and Bob share is noisy. We assume that Eve believes the channel to be noisier than it is, which allows Alice and Bob to communicate at a linear rate in the number of channel uses. This assumption is not unreasonable, especially when Alice and Bob have been systemetically deceiving Eve by adding extra noise. We also assume that Alice and Bob are using an error-correcting code powerful enough to correct errors induced by the channel Eve believes to be connecting them, or else she would become suspicious. Eve would also become suspicious if the pattern of errors Alice uses to encode her secret information does not match the typical errors induced by the channel that Eve expects. 

In Section \ref{achiev} we formalize our notion of quantum steganography where secret messages are hidden in the syndromes of an error-correcting code. We outline a specific steganographic encoding where Alice is able to emulate a bit flip or depolarizing channel $\mathcal{N}_{p+\delta p}$ (the channel Eve believes to be connecting them) on her encoded secret message and covertext, where the actual physical channel is $\mathcal{N}_{p}$.   We also calculate the amount of key consumed in our protocol. In Section \ref{converse} we prove upper bounds on the amount of steganographic communication possible over these channels, and show that these bounds are asymptotically equal to the rates achieved in the previous section. Finally, in Section \ref{conc} we summarize our results, and discuss quantum steganography for general quantum channels, conjecturing a capacity formula for general quantum steganographic communication.

\section{Achievability}\label{achiev}
\subsection{Bit Flip Channel}\label{bitflipa}
Suppose that Alice wishes to communicate steganographically to Bob by secretly sending him a message $m$ drawn from a set of possible messages $\mathcal{M}$, assumed to all be equally likely. Alice and Bob can communicate via a quantum channel, but the eavesdropper Eve can monitor their communications over this channel if she chooses. They cannot communicate clasically without Eve intercepting their communications, but before the protocol began they exchanged a secret key, in the form of an arbitrarily long string of random bits, unknown to Eve. We assume that Eve believes the quantum channel shared between Alice and Bob to be a bit flip channel with error rate $p+\delta p$, i.e.,
\begin{equation}\label{eq:bitflip}
\mathcal{N}^{BF}_{p+\delta p}(\rho)=(1-(p+\delta p))\rho+(p+\delta p)X\rho X.
\end{equation}
However, the actual physical channel between Alice and Bob is $\mathcal{N}^{BF}_{p}$. First, Alice encodes an innocent state, i.e. the covertext $\rho_{c}$, into a nondegenerate quantum error-correcting code (QECC) on $N$ qubits. This code should be able to correct typical errors induced by the channel $(\mathcal{N}^{BF}_{p+\delta p})^{\otimes N}$. Next, depending on the secret key $k\in\mathcal{K}$ and the message $m\in\mathcal{M}$ that she would like to send, she applies the error 
\begin{equation}\label{eq:codewords}
E^{N}(k,m)=E_{1}(k,m)\otimes...\otimes E_{N}(k,m)
\end{equation}
to her state. This produces the codeword corresponding to the message $m$. If her message to Bob is a quantum state, she can prepare the system in a superposition of these codewords. These codewords are generated by applying errors drawn randomly from the channel $(\mathcal{N}_{q}^{BF})^{\otimes N}$, using the shared secret key $k$ as the source of randomness. That is, the errors $X$ or $I$ on each qubit are drawn from the product distribution $p_{E^{N}}(e^{N})$, where $p_{E}(e)$ is given by
\begin{align}\label{eq:randomcodeprob}
&p_{E}(X)=q, \nonumber \\
 &p_{E}(I)=1-q,
\end{align}
where $q=\delta p/(1-2p)$. Since the set of errors is selected using the shared secret key $k$, Bob knows which codeword corresponds to each message $m$.

The errors given by Eq.~\eqref{eq:codewords} are typical errors associated with the channel $(\mathcal{N}_{q}^{BF})^{\otimes N}$ \cite{klesse2007approximate, klesse2008random}.  By the asymptotic equipartition theorem \cite{wilde2013quantum}, for large enough $N$, it is highly likely that each of these codewords that Alice generates is a typical sequence with a sample entropy close to $H(E)=-(1-q)\log (1-q)-q\log q=h(q)$. Furthermore, it follows from a simple calculation that $\mathcal{N}_{p}\circ\mathcal{N}_{q}=\mathcal{N}_{p+\delta p}$ if we set $q=\delta p/(1-2p)$. This will become important later when we discuss the secrecy of this protocol. 

 Alice then sends her state through the channel $(\mathcal{N}_{p}^{BF})^{\otimes N}$. We are now essentially in the scenario of classical random coding over a classical bit-flip channel with parameter $p$. By the asymptotic equipartition theorem for conditionally typical sequences \cite{wilde2013quantum}, for each input sequence (i.e., error $E^{N}(k,m)$ applied to the encoded covertext) there is a corresponding conditionally typical set of errors $\{F^{N}(k,m)\}$ which has the following properties: its total probability is close to 1, its size is $\approx 2^{nH(F|E)}$, and the probability of each conditionally typical error given knowledge of the input error $E^{N}(k,m)$ is $\approx 2^{-nH(F|E)}$. 

With high probability, the error $F^{N}$ Bob observes  will be a typical error of the channel $(\mathcal{N}_{p+\delta p}^{BF})^{\otimes N}$. We know from Shannon's noisy channel coding theorem that if Alice and Bob set the number of messages $|\mathcal{M}|=2^{NR}$ such that
\begin{equation}
2^{NR}\approx \frac{2^{NH(F)}}{2^{NH(F|E)}}=2^{N(H(F)-H(F|E))},
\end{equation}
then Bob is able to decode correctly with high probability \cite{cover2012elements, nielsen2010quantum, wilde2013quantum}  which error $E^{N}(k,m)$ was applied by Alice, as long as the code is nondegenerate. For our protocol, it is straightforward to calculate that $H(F)=h(p+q-2pq)=h(p+\delta p)$ for $q=\delta p/(1-2p)$, and $H(F|E)=h(p)$. Hence Alice can communicate
\begin{equation}\label{eq:bfachiev}
M=\log|\mathcal{M}|\approx N(h(p+\delta p)-h(p))
\end{equation}
bits of information to Bob steganographically.

Moreover, this protocol does not arouse suspicion from Eve. We say that this protocol is $\textit{secret}$, because the state passing through the channel is to good approximation the state Eve would expect to see. To see this, note that
\begin{align}\label{eq:secrecyderivation}
&\sum_{k\in\mathcal{K}}\sum_{m\in \mathcal{M}}p_{k}(\mathcal{N}^{BF}_{p})^{\otimes N}(E^{N}(k,m)V\rho_{c}V^{\dagger}E^{N}(k,m)) \nonumber \\
&=(\mathcal{N}^{BF}_{p})^{\otimes N}(\sum_{k\in\mathcal{K}}\sum_{m\in\mathcal{M}}p_{k}E^{N}(k,m)V\rho_{c}V^{\dagger}E^{N}(k,m))\nonumber \\
&\approx (\mathcal{N}_{p}^{BF}\circ\mathcal{N}_{\delta p/(1-2p)}^{BF})^{\otimes N}(V\rho_{c}V^{\dagger}) \nonumber \\
&=(\mathcal{N}_{p+\delta p}^{BF})^{\otimes N}(V\rho_{c}V^{\dagger}),
\end{align}
where $V$ is the isometry corresponding to the QECC Alice and Bob are using. The first equality follows from linearity of quantum operations. The approximate equality follows from the fact that when we average the transmitted codeword over the key and all possible messages, we are applying all the typical errors of the channel $(\mathcal{N}_{q}^{BF})^{\otimes N}$ with their correct probabilities, and hence to good approximation \cite{klesse2007approximate, klesse2008random} we are simply applying the full channel. The final equality follows from calculating the composition of these quantum operations. This is exactly the state Eve expects to observe, hence our steganographic protocol is secret to an arbitrarily good approximation.

As described above, this protocol allows Alice to transmit a classical message $m$ secretly to Bob. But in fact, by making this protocol coherent Alice can equally well transmit a quantum state---that is, a superposition of possible messages $m$. So we see that this protocol can transmit either classical or quantum information at the same rate $h(p+\delta p)-h(p)$. The one significant difference between these two cases is that if Eve actually carries out a measurement of the error on the transmitted state, this would destroy the superpositions of a quantum message, but not affect the ability to transmit classical messages. So this protocol works for secret quantum communication if it is assumed that Eve only sometimes checks the code blocks transmitted from Alice to Bob. As we did in the case of steganographic communication over a noiseless channel \cite{sutherland2018quantum}, we can show this by considering a protocol in which Alice sends a subsystem $M$ to Bob which is maximally entangled with a reference subsystem $R$ (see Figure \ref{fig:protocol}).
\subsection{Depolarizing Channel}\label{dca}
Suppose that Eve believes the quantum channel shared between Alice and Bob is a depolarizing channel with error rate $p+\delta p$, i.e.,
\begin{align}\label{eq:depolarizingchannel}
\mathcal{N}^{DC}_{p+\delta p}(\rho)=(&1-(p+\delta p))\rho\nonumber\\
&+\frac{p+\delta p}{3}(X\rho X+Y\rho Y+Z\rho Z),
\end{align}
where the actual physical channel between Alice and Bob is $\mathcal{N}^{DC}_{p}$. The protocol for Alice and Bob to communicate steganographically in this scenario is nearly identical to the protocol described in the previous subsection. First, Alice encodes an innocent state, i.e,. the covertext $\rho_{c}$, into a nondegenerate quantum error-correcting code (QECC) on $N$ qubits. This code should be able to correct typical errors induced by the channel $(\mathcal{N}^{DC}_{p+\delta p})^{\otimes N}$. Next, depending on the secret key $k\in\mathcal{K}$ and the message $m\in\mathcal{M}$ that she would like to send, she applies the error 
\begin{equation}\label{eq:dcerrors}
G^{N}(k,m)=G_{1}(k,m)\otimes...\otimes G_{N}(k,m)
\end{equation}
to her state. This produces the codeword corresponding to the message $m$. If her message to Bob is a quantum state, she can prepare the system in a superposition of these codewords. These codewords are generated by applying errors drawn randomly from the channel $(\mathcal{N}_{q}^{DC})^{\otimes N}$, using the shared secret key $k$ as the source of randomness. That is, the errors $X$, $Y$, $Z$ or $I$ on each qubit are drawn from the product distribution $p_{G^{N}}(g^{N})$, where $p_{G}(g)$ is given by
\begin{align}\label{eq:randomcodeprobdc}
&p_{G}(X)=p_{G}(Y)=p_{G}(Z)=q/3, \nonumber \\
 &p_{G}(I)=1-q,
\end{align}
and $q=\delta p/(1-4p/3)$. Since the errors are selected using the shared secret key $k$, Bob knows what codeword corresponds to each message $m$.

The errors given by Eq.~\eqref{eq:dcerrors} are typical errors associated with the channel $(\mathcal{N}_{q}^{DC})^{\otimes N}$ \cite{klesse2007approximate, klesse2008random}.  By the asymptotic equipartition theorem \cite{wilde2013quantum}, for large enough $N$, it is highly likely that each of these codewords that Alice generates is a typical sequence with a sample entropy close to $H(G)=-(1-q)\log (1-q)-q\log(q/3)\equiv s(q)$, where $s(q)$ is the classical Shannon entropy of the depolarizing channel on one qubit with error parameter $q$ in the Pauli representation. Furthermore, it follows from a simple calculation that $\mathcal{N}_{p}\circ\mathcal{N}_{q}=\mathcal{N}_{p+\delta p}$ if we set $q=\delta p/(1-4p/3)$, which is important for secrecy as discussed in the previous subsection.

 Alice then sends her state through the channel $(\mathcal{N}_{p}^{DC})^{\otimes N}$. Following the same random coding argument described for the bit-flip channel, with high probability, the error $J^{N}$ Bob observes  will be a typical error of the channel $(\mathcal{N}_{p+\delta p}^{DC})^{\otimes N}$. We know from Shannon's noisy channel coding theorem that if Alice and Bob set the number of messages $|\mathcal{M}|=2^{NR}$ such that
\begin{equation}
2^{NR}\approx \frac{2^{NH(J)}}{2^{NH(J|G)}}=2^{N(H(J)-H(J|G))},
\end{equation}
then Bob is able to decode correctly with high probability \cite{cover2012elements, nielsen2010quantum, wilde2013quantum}  which error $G^{N}(k,m)$ was applied by Alice, as long as the code is nondegenerate. For our protocol, it is straightforward to calculate that $H(J)=s(p+q-4qp/3)=s(p+\delta p)$ for $q=\delta p/(1-4p/3)$, and $H(J|G)=s(p)$. Hence Alice can communicate
\begin{equation}\label{eq:dcachiev}
M=\log|\mathcal{M}|\approx N(s(p+\delta p)-s(p))
\end{equation}
classical or quantum bits of information to Bob steganographically. The proof of secrecy of this protocol is nearly identical to the one given in Eq.~\eqref{eq:secrecyderivation}.

Note that the assumption of a nondegenerate code is quite natural in the case of the bit-flip channel, which is essentially classical; but not as much so for the depolarizing channel, where the errors do not commute.  We believe that this general procedure for encoding will work for degenerate codes as well, but the achievable rate may be lower, and will require an analysis specific to the code in question.  We will return to this point at the end of the paper, where we conjecture a general formula for the steganographic rate of a quantum channel using general quantum codes.

\subsection{Secret key consumption}
Here we analyze how much secret key is used by the encodings outlined above. The details of the encoding---that is, how each message $m$ is mapped to a codeword for a particular key element $k$---we assume have been decided between Alice and Bob ahead of time. Therefore secret key is required to pick the subsets of errors used in the encoding, but it is not needed otherwise.

Before the protocol begins, Alice and Bob divide the set of typical errors of the channel $(\mathcal{N}^{BF}_{\delta p/(1-2p)})^{\otimes N}$ into $n$ nonoverlapping subsets of size $|\mathcal{M}|=2^{N(h(p+\delta p)-h(p))}$ each, where
\begin{equation}
n=\frac{2^{Nh(\delta p/(1-2p))}}{2^{N(h(p+\delta p)-h(p))}}.
\end{equation}
For each transmitted block, Alice and Bob must randomly choose one of these $n$ subsets to encode her messages. This requires a number of bits $K$ of secret key, 
\begin{equation}
K=\log_{2}n=N(h(\delta p/(1-2p))-h(p+\delta p)+h(p)),
\end{equation}
which is positive for $p+\delta p <0.5$. Therefore the key consumption scales linearly with $N$. Notice in the limit where the physical channel is noiseless i.e., $p=0$, we have that $K=0$, which agrees with our result in \cite{sutherland2018quantum} where it was shown that only a sublinear amount of key is needed for encoding across noiseless channels.

As discussed in \cite{sutherland2018quantum}, using this amount $K$ of shared secret is key is sufficient to make the steganographic protocol secret, but not necessarily secure. That is, Eve should not become suspicious if she observes the state passing through the channel. However, if for some reason she knew a message was being sent, she would be able to deduce significant information about the message.

This can be prevented by first encrypting the message before doing the steganographic encoding. If we wish to make the protocol both secret and secure, encryption would require $M$ bits of secret key in the case of an $M$-bit classical message (using a one-time pad), or $2M$ bits of secret key in the case of an $M$-qubit quantum message (by twirling). Thus the rate of key consumption would be increased by $R$ (classical) or $2R$ (quantum, where $R$ is the steganographic rate).
\section{Secrecy, Reliability, and Bounds}\label{converse}
\begin{figure}
\includegraphics[width=0.5\textwidth,height=0.5\textheight,keepaspectratio]{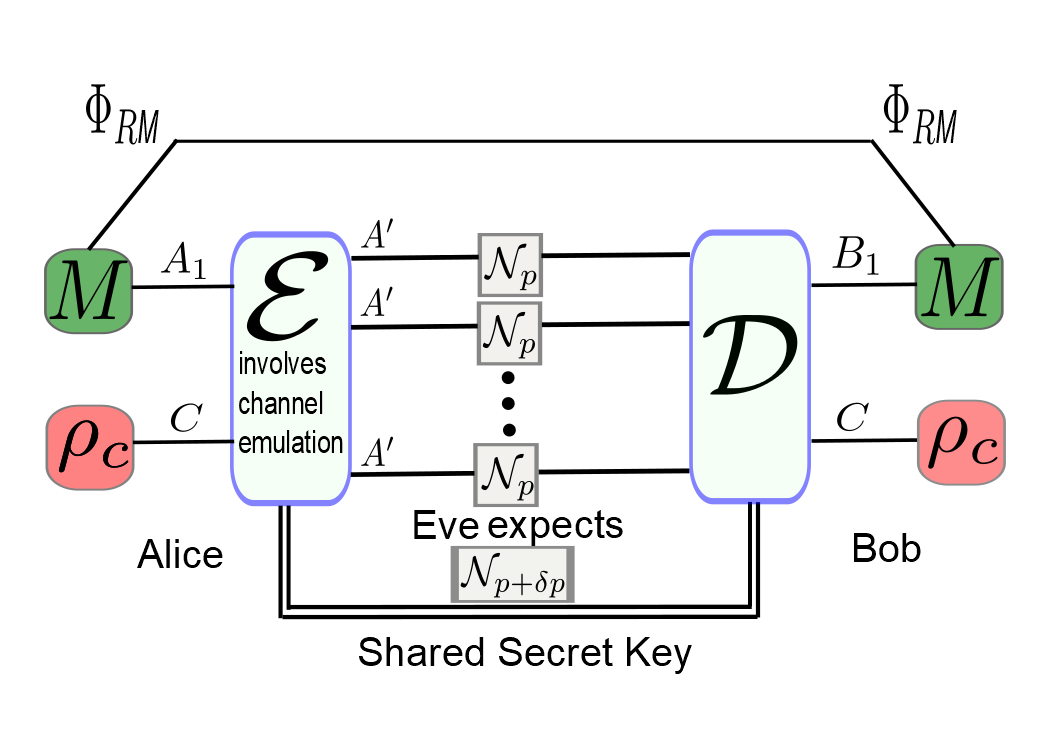}
\caption{The information processing task we consider, of Alice sending $M$ stego qubits to Bob over a quantum channel $\mathcal{N}_{p}$ (which is either the quantum bit flip channel or depolarizing channel), which Eve believes to be noisier. Eve's ignorance of the channel is characerized by the parameter $\delta p$. Dependent on the secret key $k$, Alice encodes her message subsystem $M$ and an innocent covertext $\rho_c$ into a suitable quantum error-correcting code in such a way that once passed through the physical channel, it looks as though typical errors of the channel $\mathcal{N}_{p+\delta p}$ have been applied. Bob then decodes the message and covertext using his copy of the shared secret key $k$. Alice's message is entangled with a reference system $R$. The ability to transmit entanglement can be used to bound the ability to do general quantum communication.}
\label{fig:protocol}
\end{figure}
\subsection{The information processing task}
Now we wish to put a bound on the amount of information that can be sent with the steganographic scenario outlined above. Recall that Alice is using Eve's ignorance of the actual noise rate of the physical channel to hide her message. We will consider the quantum information processing task known as \textit{entanglement transmission} (visualized in Figure \ref{fig:protocol}) in this section. The maximum rate of entanglement transmission is clearly an upper bound on the maximum rate of quantum communication and so we will use this to derive a bound on the steganographic transmission rate. Alice prepares a secret message of $M=\log_{2}|A_{1}|$ qubits along with an innocent covertext $\rho_{c}$. Her secret message qubits are maximally entangled with a reference system R. Her covertext will be encoded into the $N$-qubit quantum error-correcting code. Therefore her encoded state, which is dependent on the secret key element k, can be written as:
\begin{equation}
\omega_{k,A^{'n}R}\equiv\mathcal{E}_{k,A_{1}C\rightarrow A^{'n}}(\rho_{c}\otimes\Phi_{A_{1}R}).
\end{equation}
The dependence of the encoding on the secret key corresponds to choosing among the different sets of typical errors of the channel $\mathcal{N}_{q}$ in the protocols from the previous section. To someone (like Eve) who does not know the secret key $k$, the state is effectively
\begin{equation}
\sum_{k}p_{k}\mathcal{N}_{p}^{\otimes N}(\omega_{k,A^{'n}R})=\mathcal{N}_{p}^{\otimes N}(\omega_{A^{'n}R}),
\end{equation}
where we have used linearity of quantum operations and $\omega_{A^{'n}R}\equiv\sum_{k}p_{k}\omega_{k,A^{'n}R}$ is the state averaged over all possible values of the secret key $k$ with probabilities $p_{k}$. (We can choose this probability to be uniform for simplicity, if we so desire.)

What is a good way to guarentee secrecy from Eve? We propose the following \textit{secrecy} condition:
\begin{equation}\label{eq:secrecy}
\frac{1}{2}||\mathcal{N}_{p}^{\otimes N}(\Tr_{R}[\omega_{A^{'n}R}])-\mathcal{N}^{\otimes N}_{p+\delta p}(V\rho_{c}V^{\dagger})||\leq\delta,
\end{equation}
where $\mathcal{N}_{p+\delta p}$ is what Eve believes the physical channel to be, $V$ is an isometry representing the encoding of the covertext into a suitably chosen codeword (one which can correct typical errors induced by the channel $\mathcal{N}_{p+\delta p}$) and $\delta>0$ is some small parameter. This condition means that if Eve observes the quantum state, it will be effectively indistinguishable from an encoded covertext being sent through the noisy quantum channel $\mathcal{N}_{p+\delta p}$.

It is also important to discuss the requirement of \textit{recoverability}. When Bob receives the state, he applies his decoder $\mathcal{D}_{k,A^{'n}\rightarrow B_{1}C}$ to obtain the original state $\rho_{c}\otimes\Phi_{B_{1}R}$. The recoverability condition can be written as follows:
\begin{equation}\label{eq:recoverability}
\frac{1}{2}||\mathcal{D}_{k,A^{'n}\rightarrow B_{1}C}(\mathcal{N}_{p}^{\otimes N}\otimes I_{R}(\omega_{k,A^{'n}R}))-\rho_{c}\otimes\Phi_{B_{1}R}||_{1}\leq \epsilon
\end{equation}
for all $k$, where $\epsilon>0$ is a small parameter.

\subsection{Upper bound on steganographic rate}
We are now in a position to put a bound on the number of qubits $M$ that can be sent reliably and steganographically from Alice to Bob. First we define $\sigma_{E}=\mathcal{N}_{p+\delta p}^{\otimes N}(V\rho_{c}V^{\dagger})$ and apply the Fannes-Audeneart inequality \cite{audenaert2007sharp} to the secrecy condition in Eq.~\eqref{eq:secrecy}:
\begin{equation}\label{eq:secretineq}
H\big(\mathcal{N}_{p}^{\otimes N}(\Tr_{R}[\omega_{A^{'n}R}])\big)\leq H(\sigma_{E})+g(N,\delta)
\end{equation}
where $g(N,\delta)\equiv\delta N+h_{2}(\delta)$, and $h_{2}(\cdot)$ is the binary entropy function. Also, from the recoverability condition we have
\begin{align}\label{eq:Mineq}
M&=\log|A_{1}|=I(R\rangle B_{1})_{\Phi_{B_{1}R}}\nonumber \\
&\leq I(R\rangle B_{1})_{D_{k}(\mathcal{N}_{p}^{\otimes N}(\omega_{k}))} +\epsilon N+(1+\epsilon)h_{2}(\epsilon/[1+\epsilon]) \nonumber \\
&\leq I(R\rangle A^{'n})_{\mathcal{N}_{p}^{\otimes N}(\omega_{k})}+f(N,\epsilon) \nonumber \\
&=H\big(\mathcal{N}_{p}^{\otimes N}(\Tr_{R}[\omega_{k,A^{'n}R}])\big)\nonumber \\
&\hspace{13mm}-H\big(\mathcal{N}_{p}^{\otimes N}\otimes I_{R}(\omega_{k,A^{'n}R})\big)+f(N,\epsilon),
\end{align}
where $f(N,\epsilon)\equiv \epsilon N+(1+\epsilon)h_{2}(\epsilon/[1+\epsilon])$. The first equality follows from the fact that the coherent information of a maximally entangled state is just the logarithm of the dimension of one of the subsystems. The first inequality follows from the Alicki-Fannes-Audeneart inequality \cite{alicki2004continuity} applied to the recoverability condition given in Eq.~\eqref{eq:recoverability}. The second inequality is a quantum data processing inequality \cite{wilde2013quantum}. The last equality follows from the definition of the coherent information.

Furthermore, using the concavity of the von Neumann entropy and linearity of quantum operations we have that
\begin{align}
&\min_{k\in\mathcal{K}}H\big(\mathcal{N}_{p}^{\otimes N}(\Tr_{R}[\omega_{k,A^{'n}R}])\big)\nonumber \\
&\leq \sum_{k}p_{k}H\big(\mathcal{N}_{p}^{\otimes N}(\Tr_{R}[\omega_{k,A^{'n}R}])\big)\nonumber \\
&\leq H\big(\mathcal{N}_{p}^{\otimes N}(\Tr_{R}[\sum_{k}p_{k}\omega_{k,A^{'n}R}])\big)\nonumber \\
&=H\big(\mathcal{N}_{p}^{\otimes N}(\Tr_{R}[\omega_{A^{'n}R}])\big),
\end{align}
and for many cases we expect $H(\mathcal{N}_{p}^{\otimes N}(\Tr_{R}[\omega_{k,A^{'n}R}]))$ to be roughly the  same for every $k$ (see Sec.~\ref{bfconv} and ~\ref{dcconv}). Thus 
\begin{equation}\label{eq:concavineq}
H\big(\mathcal{N}_{p}^{\otimes N}(\Tr_{R}[\omega_{k,A^{'n}R}])\big)\leq H\big(\mathcal{N}_{p}^{\otimes N}(\Tr_{R}[\omega_{A^{'n}R}])\big)
\end{equation}
for all $k$. Now putting Eq.~\eqref{eq:secretineq}, \eqref{eq:Mineq}, and \eqref{eq:concavineq}  together we arrive at our main result for this section, which states that Alice can secretly and reliably send $M$ stego qubits to Bob, where $M$ is bounded above by
\begin{align}\label{eq:ratebound}
M&\leq H(\sigma_{E})\nonumber \\
&-H\big((\mathcal{N}_{p}^{\otimes N}\otimes I_{R})(\omega_{k,A^{'n}R})\big)+g(N,\delta)+f(N,\epsilon).
\end{align}
Thus, if we can compute a maximum for $H(\sigma_{E})=H(\mathcal{N}_{p+\delta p}^{\otimes N}(\rho))$ where $\rho$ is pure (because $V$ is an isometric encoding and $\rho_{c}$ is pure), and also compute a lower bound for $H\big((\mathcal{N}_{p}^{\otimes N}\otimes I_{R})(\omega_{k,A^{'n}R})\big)$ (or compute it explictly, recalling that $\omega_{k,A^{'n}R}$ is a pure state), then we have a tight upper bound on the number of qubits $M$ that can be sent steganographically over a noisy quantum channel $\mathcal{N}_{p}$. 
\subsection{Upper bounds for specific channels}
For the channels discussed in the achievability section of this paper, we can now apply our result given in Eq.~\eqref{eq:ratebound}, where we make the implicit assumption that Alice is using a nondegenerate code. Though our result given by Eq.~\eqref{eq:ratebound} is true in general, for a degenerate code the number of distinct error syndromes is smaller (depending on the code), and the bounds discussed here and achievable rates discussed in the previous section would be adjusted.
\subsubsection{The bit flip channel}\label{bfconv}
For the bit flip channel with parameter $p+\delta p$ given by Eq.~\eqref{eq:bitflip}, the maximum of $H((\mathcal{N}^{BF}_{p+\delta p})^{\otimes N}(\rho))$ over all $N$-qubit pure states $\rho$ is $Nh(p+\delta p)$ where $h(p+\delta p)=-(p+\delta p)\log(p+\delta p)-(1-(p+\delta p))\log(1-(p+\delta p))$ is the entropy of a single qubit sent through this bit flip channel. To prove this, consider some pure state $\rho=\ket{\psi}\bra{\psi}$. Then
\begin{equation}
(\mathcal{N}^{BF}_{p+\delta p})^{\otimes N}(\ket{\psi}\bra{\psi})=\sum_{s}p(s)X^{s}\ket{\psi}\bra{\psi}X^{s}
\end{equation}
where we are summing over all binary strings $s$ of length $N$; $X^{s}$ is the operator acting on $N$ qubits with an $X$ acting at every location where $s$ has a 1 and an $I$ where $s$ has a 0. The probability $p(s)$ is given by
\begin{equation}
p(s)=p^{w(s)}(1-p)^{N-w(s)},
\end{equation}
where $w(s)$ is the weight of string $s$. The Shannon entropy of this distribution is $Nh(p+\delta p)$ since it is a binomial distribution. The von Neumann entropy is the minimum Shannon entropy over all possible ensemble decompositions of the given state, and it is not hard to check that it is achieved when $\ket{\psi}$ is a $Z$ eigenstate. Thus we have computed a maximum for the first term on the right hand side of Eq.~\eqref{eq:ratebound}.

Now we compute a lower bound for the second term on the right hand side of Eq.~\eqref{eq:ratebound} in the case of the bit flip channel, i.e., $H((\mathcal{N}^{BF}_{p})^{\otimes N}\otimes I_{R})(\omega_{k,A^{'n}R}))$. First note that we can write 
\begin{align}\label{eq:typatyp}
&((\mathcal{N}_{p}^{BF})^{\otimes N}\otimes I_{R})(\omega_{k,A^{'n}R}) \nonumber \\
&=\sum_{\underline{i}\in\mathcal{T}}(E_{\underline{i}}\otimes I_{R})\omega_{k,A^{'n}R}(E_{\underline{i}}^{\dagger}\otimes I_{R})\nonumber \\
&+\sum_{\underline{i}\notin\mathcal{T}}(E_{\underline{i}}\otimes I_{R})\omega_{k,A^{'n}R}(E^{\dagger}_{\underline{i}}\otimes I_{R}),
\end{align}
where the index is $\underline{i}=i_{1}i_{2}...i_{N}$, the errors $E_{\underline{i}}$ are given by
\begin{equation}
E_{\underline{i}}=A_{i_{1}}\otimes...\otimes A_{i_{N}},
\end{equation}
and $\mathcal{T}$ is the set of typical sequences $\underline{i}$ corresponding to the typical errors of the channel $(\mathcal{N}^{BF}_{p})^{\otimes N}$ \cite{klesse2007approximate, klesse2008random}.

Recall we are making the assumption that the QECC Alice is using to correct the typical errors of the channel $(\mathcal{N}_{p+\delta p}^{BF})^{\otimes N}$ is nondegenerate. Because of this, we can infer that for each $k$, her encoded state $\omega_{k,A^{'n}R}$ forms a nondegenerate code for the channel $(\mathcal{N}_{p}^{BF})^{\otimes N}$. This follows from the discussion in Section \ref{bitflipa}. This means that on a valid codeword in the QECC, for $\underline{i}\in\mathcal{T}$ the typical errors $E_{\underline{i}}$ all have distinct error syndromes, and act as unitaries that move the code space to a distinct, orthogonal subspace labeled by $\underline{i}$. So an error $E_{\underline{i}}$ occurs with a fixed probability $p_{\underline{i}}$ for all valid codewords of the QECC. Recall also that since these errors are typical, they have almost all the probability, i.e. 
\begin{equation}\label{eq:typicalproperty}
\sum_{i\in\mathcal{T}}p_{i}=1-\epsilon
\end{equation}
for arbitrarily small $\epsilon>0$ (in the limit of large $N$). From Eq.~\eqref{eq:typatyp} we have that
\begin{align}
&H\big(((\mathcal{N}_{p}^{BF})^{\otimes N}\otimes I_{R})(\omega_{k,A^{'n}R})\big)\nonumber \\
&\geq (1-\epsilon)H\big(\frac{1}{1-\epsilon}\sum_{\underline{i}\in\mathcal{T}}(E_{\underline{i}}\otimes I_{R})\omega_{k,A^{'n}R}(E_{\underline{i}}^{\dagger}\otimes I_{R})\big)\nonumber \\
&+\epsilon H\big(\frac{1}{\epsilon}\sum_{\underline{i}\notin\mathcal{T}}(E_{\underline{i}}\otimes I_{R})\omega_{k,A^{'n}R}(E^{\dagger}_{\underline{i}}\otimes I_{R})\big)\nonumber \\
&\geq (1-\epsilon)H\big(\frac{1}{1-\epsilon}\sum_{\underline{i}\in\mathcal{T}}(E_{\underline{i}}\otimes I_{R})\omega_{k,A^{'n}R}(E_{\underline{i}}^{\dagger}\otimes I_{R})\big)
\end{align}
where the first inequality follows from the concavity of the von Neumann entropy and Eq.~\eqref{eq:typicalproperty} , and the second inequality follows because the term proportional to $\epsilon$ is positive. Continuing, we have
\begin{align}
&(1-\epsilon)H\big(\frac{1}{1-\epsilon}\sum_{\underline{i}\in\mathcal{T}}(E_{\underline{i}}\otimes I_{R})\omega_{k,A^{'n}R}(E_{\underline{i}}^{\dagger}\otimes I_{R})\big)\nonumber\\
&=-\sum_{\underline{i}\in\mathcal{T}}p_{\underline{i}}\log\frac{p_{\underline{i}}}{1-\epsilon}=\sum_{i\in\mathcal{T}}\big(p_{\underline{i}}\log(1-\epsilon)-p_{\underline{i}}\log p_{\underline{i}}\big)\nonumber\\
&=(1-\epsilon)\log(1-\epsilon)-\sum_{\underline{i}\in\mathcal{T}}p_{\underline{i}}\log p_{\underline{i}}\nonumber \\
&\geq Nh(p)-\mathcal{O}(\epsilon).
\end{align}
The first equality follows from the definition of the von Neumann entropy and Eq.~\eqref{eq:typatyp}. The inequality follows from performing a Taylor expansion on $\log(1-\epsilon)$ and from the fact that for the bit flip channel $(\mathcal{N}_{p}^{BF})^{\otimes N}$:
\begin{align}
-\sum_{\underline{i}\in\mathcal{T}}p_{\underline{i}}\log p_{\underline{i}}&\approx -\sum_{\underline{i}\in\mathcal{T}}2^{-Nh(p)}\log 2^{-Nh(p)}\nonumber\\
&=Nh(p)\sum_{\underline{i}\in\mathcal{T}}2^{-Nh(p)}\nonumber\\
&\approx Nh(p)2^{Nh(p)}2^{-Nh(p)}=Nh(p),
\end{align}
where the approximate equalities follow directly from the theory of typical sequences. Thus 
\begin{equation}
H\big(((\mathcal{N}_{p}^{BF})^{\otimes N}\otimes I_{R})(\omega_{k,A^{'n}R})\big)\geq Nh(p)-\mathcal{O}(\epsilon),
\end{equation}
and combining this with the upper bound we computed for the first term in Eq.~\eqref{eq:ratebound} gives
\begin{equation}
M\leq N(h(p+\delta p)-h(p))
\end{equation}
for a sufficiently reliable and secret protocol and large enough block size $N$. Comparing this to Eq.~\eqref{eq:bfachiev}, we have that the encoding described in the previous section for steganography over the channel $\mathcal{N}_{p}^{BF}$ where Eve expects the channel to be $\mathcal{N}_{p+\delta p}^{BF}$ is essentially optimal.
\subsubsection{The depolarizing channel}\label{dcconv}
Unfortunately, for the depolarizing channel $\mathcal{N}^{DC}$ we do not know what $N$-qubit pure state $\rho$ maximizes $H((\mathcal{N}^{DC}_{p+\delta p})^{\otimes N}(\rho))$. However, we can still bound this quantity, i.e., give an upper bound on the first term in Eq.~\eqref{eq:ratebound}.  Consider the action of this channel on an $N$ qubit pure state as follows:
\begin{equation}
(\mathcal{N}^{DC}_{p+\delta p})^{\otimes N}(\rho)\approx\sum_{j}E_{j}\rho E_{j}^{\dagger}
\end{equation}
where $\{E_{j}\}$ is the set of typical errors associated with $N$ applications of the channel $(\mathcal{N}_{p+\delta p}^{DC})^{\otimes N}$. Recall that we are choosing our isometric encoding to correct for typical errors of the channel Eve beleives to be connecting Alice and Bob, i.e. $\mathcal{N}^{DC}_{p+\delta p}$. Furthermore, we're assuming that our code is nondegenerate. Therefore  the states $E_{j}\rho E_{j}^{\dagger}$ are all orthogonal to each other for $\rho$ in the codespace, and $\Tr[E_{j}\rho E_{j}^{\dagger}]=p_{j}$ where $p_{j}$ are the typical probabilities associated with the errors $E_{j}$. The von Neumann entropy is the Shannon entropy minimized over all possible decompositions, so the entropy of this state is clearly
\begin{align}\label{eq:dcupperbound}
H(\sigma_{E})&=H((\mathcal{N}_{p+\delta p}^{DC})^{\otimes N}(V\rho_{c}V^{\dagger}))\nonumber\\
&\leq -\sum_{j}p_{j}\log(p_{j})\nonumber\\
&=-\sum_{j}2^{-Ns(p)}\log 2^{-Ns(p)}\nonumber\\
&\approx Ns(p)2^{Ns(p)}2^{-Ns(p)}=Ns(p)
\end{align}
In a similar argument to the one given in Sec.~\ref{bfconv} , we can give a lower bound for the second term on the right hand side of Eq.~\eqref{eq:ratebound} in the case of the depolarizing channel. We get that
\begin{equation}
H(((\mathcal{N}_{p}^{DC})^{\otimes N}\otimes I_{R})(\omega_{k,A^{'n}R}))\geq Ns(p)-\mathcal{O}(\epsilon),
\end{equation}
where $\epsilon$ becomes arbitrarily small for large $N$. Combining this with the upper bound given in Eq.~\eqref{eq:dcupperbound} we have that
\begin{equation}
M\leq N(s(p+\delta p)-s(p))
\end{equation}
for a sufficiently reliable and secret protocol and large enough block size $N$. Comparing this to Eq.~\eqref{eq:dcachiev} we have that the encoding described in the previous section for steganography over the channel $\mathcal{N}_{p}^{DC}$ where Eve expects the channel to be $\mathcal{N}_{p+\delta p}^{DC}$ is essentially optimal, at least when we restrict ourselves to nondegenerate codes.
\section{Conclusion}\label{conc}
Characterizing secret communication over noisy quantum channels is an interesting problem from both a practical and theoretical perspective. Here we have shown that two parties are able to communicate secretly with each other  at a nonzero rate over a bit-flip or a depolarizing channel $\mathcal{N}_{p}$ using a shared secret key, without arousing suspicion from a potential eavesdropper Eve, so long as Eve believes the channel to be noisier than it really is. Eve can be made to believe this through Alice and Bob systematically adding extra noise to the channel prior to secret communication. In this paper we gave explicit bounds on the number of stego qubits that Alice can send to Bob by hiding her secret message in the syndromes of a nondegenerate error-correcting code designed to correct the typical errors of the channel Eve believes: $\mathcal{N}_{p+\delta p}$. We also gave explicit encodings that achieve these bounds.

Interesting future work should include a generalization of these results to steganography over general quantum channels $\mathcal{N}$. It is possible that in order to achieve the maximum possible rates in this scenario that degenerate codes are needed. For example, it is likely that the steganographic capacity we calculated for the depolarizing channel in this paper could be improved in this way. It is also possible that coding across multiple codeblocks using degenerate quantum codes could increase the steganographic capacity.

If the actual physical channel shared between Alice and Bob is $\mathcal{N}$, and the channel Eve believes is $\mathcal{M}$, then what is the quantum steganographic capacity? In this paper we proved that for the bit-flip channel, the rate is the difference of quantum capacities, i.e., $Q(\mathcal{N}^{BF}_{p})-Q(\mathcal{N}^{BF}_{p+\delta p})=N(1-h(p)-1+h(p+\delta p))=N(h(p+\delta p)-h(p))$. Also, allowing for our restriction to nondegenerate codes, this is true for the depolarizing channel as well. We conjecture that one might be able to prove that the steganographic rate in general will be $Q(\mathcal{N})-Q(\mathcal{M})$. This will require proof methods that go beyond those of the current paper, but we believe that this wll be an area of fruitful future study.

\section*{Acknowledgments}
Thanks to Yi-Hsiang Chen and Namit Anand for helpful discussions.  This research was supported in part by  NSF Grants CCF-1421078 and QIS-1719778, and by an IBM Einstein Fellowship at the Institute for Advanced Study.
\bibliography{noisystegobib}

\begin{thebibliography}{22}%
\makeatletter
\providecommand \@ifxundefined [1]{%
 \@ifx{#1\undefined}
}%
\providecommand \@ifnum [1]{%
 \ifnum #1\expandafter \@firstoftwo
 \else \expandafter \@secondoftwo
 \fi
}%
\providecommand \@ifx [1]{%
 \ifx #1\expandafter \@firstoftwo
 \else \expandafter \@secondoftwo
 \fi
}%
\providecommand \natexlab [1]{#1}%
\providecommand \enquote  [1]{``#1''}%
\providecommand \bibnamefont  [1]{#1}%
\providecommand \bibfnamefont [1]{#1}%
\providecommand \citenamefont [1]{#1}%
\providecommand \href@noop [0]{\@secondoftwo}%
\providecommand \href [0]{\begingroup \@sanitize@url \@href}%
\providecommand \@href[1]{\@@startlink{#1}\@@href}%
\providecommand \@@href[1]{\endgroup#1\@@endlink}%
\providecommand \@sanitize@url [0]{\catcode `\\12\catcode `\$12\catcode
  `\&12\catcode `\#12\catcode `\^12\catcode `\_12\catcode `\%12\relax}%
\providecommand \@@startlink[1]{}%
\providecommand \@@endlink[0]{}%
\providecommand \url  [0]{\begingroup\@sanitize@url \@url }%
\providecommand \@url [1]{\endgroup\@href {#1}{\urlprefix }}%
\providecommand \urlprefix  [0]{URL }%
\providecommand \Eprint [0]{\href }%
\providecommand \doibase [0]{http://dx.doi.org/}%
\providecommand \selectlanguage [0]{\@gobble}%
\providecommand \bibinfo  [0]{\@secondoftwo}%
\providecommand \bibfield  [0]{\@secondoftwo}%
\providecommand \translation [1]{[#1]}%
\providecommand \BibitemOpen [0]{}%
\providecommand \bibitemStop [0]{}%
\providecommand \bibitemNoStop [0]{.\EOS\space}%
\providecommand \EOS [0]{\spacefactor3000\relax}%
\providecommand \BibitemShut  [1]{\csname bibitem#1\endcsname}%
\let\auto@bib@innerbib\@empty
%</preamble>
\bibitem [{\citenamefont {Herodotus}(1996)}]{greekstego}%
  \BibitemOpen
  \bibfield  {author} {\bibinfo {author} {\bibnamefont {Herodotus}},\
  }\href@noop {} {\emph {\bibinfo {title} {The Histories}}}\ (\bibinfo
  {publisher} {Penguin Books},\ \bibinfo {year} {1996})\BibitemShut {NoStop}%
\bibitem [{\citenamefont {Singh}(2000)}]{singh2000code}%
  \BibitemOpen
  \bibfield  {author} {\bibinfo {author} {\bibfnamefont {S.}~\bibnamefont
  {Singh}},\ }\href@noop {} {\emph {\bibinfo {title} {The code book: the secret
  history of codes and code-breaking}}}\ (\bibinfo  {publisher} {Fourth
  Estate},\ \bibinfo {year} {2000})\BibitemShut {NoStop}%
\bibitem [{\citenamefont {Petitcolas}\ \emph {et~al.}(1999)\citenamefont
  {Petitcolas}, \citenamefont {Anderson},\ and\ \citenamefont
  {Kuhn}}]{petitcolas1999information}%
  \BibitemOpen
  \bibfield  {author} {\bibinfo {author} {\bibfnamefont {F.~A.}\ \bibnamefont
  {Petitcolas}}, \bibinfo {author} {\bibfnamefont {R.~J.}\ \bibnamefont
  {Anderson}}, \ and\ \bibinfo {author} {\bibfnamefont {M.~G.}\ \bibnamefont
  {Kuhn}},\ }\href@noop {} {\bibfield  {journal} {\bibinfo  {journal}
  {Proceedings of the IEEE}\ }\textbf {\bibinfo {volume} {87}},\ \bibinfo
  {pages} {1062} (\bibinfo {year} {1999})}\BibitemShut {NoStop}%
\bibitem [{\citenamefont {Shor}(1999)}]{shor1999polynomial}%
  \BibitemOpen
  \bibfield  {author} {\bibinfo {author} {\bibfnamefont {P.~W.}\ \bibnamefont
  {Shor}},\ }\href@noop {} {\bibfield  {journal} {\bibinfo  {journal} {SIAM
  review}\ }\textbf {\bibinfo {volume} {41}},\ \bibinfo {pages} {303} (\bibinfo
  {year} {1999})}\BibitemShut {NoStop}%
\bibitem [{\citenamefont {Natori}(2006)}]{natoristego}%
  \BibitemOpen
  \bibfield  {author} {\bibinfo {author} {\bibfnamefont {S.}~\bibnamefont
  {Natori}},\ }in\ \href@noop {} {\emph {\bibinfo {booktitle} {Quantum
  Computation and Information}}}\ (\bibinfo  {publisher} {Springer},\ \bibinfo
  {year} {2006})\ pp.\ \bibinfo {pages} {235--240}\BibitemShut {NoStop}%
\bibitem [{\citenamefont {Banerjee}\ \emph {et~al.}(2012)\citenamefont
  {Banerjee}, \citenamefont {Bhattacharyya},\ and\ \citenamefont
  {Sanyal}}]{banerjeestego}%
  \BibitemOpen
  \bibfield  {author} {\bibinfo {author} {\bibfnamefont {I.}~\bibnamefont
  {Banerjee}}, \bibinfo {author} {\bibfnamefont {S.}~\bibnamefont
  {Bhattacharyya}}, \ and\ \bibinfo {author} {\bibfnamefont {G.}~\bibnamefont
  {Sanyal}},\ }\href@noop {} {\bibfield  {journal} {\bibinfo  {journal}
  {International Journal of Computer Network and Information Security}\
  }\textbf {\bibinfo {volume} {4}},\ \bibinfo {pages} {65} (\bibinfo {year}
  {2012})}\BibitemShut {NoStop}%
\bibitem [{\citenamefont {Gea-Banacloche}(2002)}]{gea2002hiding}%
  \BibitemOpen
  \bibfield  {author} {\bibinfo {author} {\bibfnamefont {J.}~\bibnamefont
  {Gea-Banacloche}},\ }\href@noop {} {\bibfield  {journal} {\bibinfo  {journal}
  {Journal of Mathematical Physics}\ }\textbf {\bibinfo {volume} {43}},\
  \bibinfo {pages} {4531} (\bibinfo {year} {2002})}\BibitemShut {NoStop}%
\bibitem [{\citenamefont {Shaw}\ and\ \citenamefont
  {Brun}(2010)}]{shaw2010hiding}%
  \BibitemOpen
  \bibfield  {author} {\bibinfo {author} {\bibfnamefont {B.~A.}\ \bibnamefont
  {Shaw}}\ and\ \bibinfo {author} {\bibfnamefont {T.~A.}\ \bibnamefont
  {Brun}},\ }\href@noop {} {\bibfield  {journal} {\bibinfo  {journal} {arXiv
  preprint arXiv:1007.0793}\ } (\bibinfo {year} {2010})}\BibitemShut {NoStop}%
\bibitem [{\citenamefont {Shaw}\ and\ \citenamefont
  {Brun}(2011)}]{shaw2011quantum}%
  \BibitemOpen
  \bibfield  {author} {\bibinfo {author} {\bibfnamefont {B.~A.}\ \bibnamefont
  {Shaw}}\ and\ \bibinfo {author} {\bibfnamefont {T.~A.}\ \bibnamefont
  {Brun}},\ }\href@noop {} {\bibfield  {journal} {\bibinfo  {journal} {Physical
  Review A}\ }\textbf {\bibinfo {volume} {83}},\ \bibinfo {pages} {022310}
  (\bibinfo {year} {2011})}\BibitemShut {NoStop}%
\bibitem [{\citenamefont {Sutherland}\ and\ \citenamefont
  {Brun}(2018)}]{sutherland2018quantum}%
  \BibitemOpen
  \bibfield  {author} {\bibinfo {author} {\bibfnamefont {C.}~\bibnamefont
  {Sutherland}}\ and\ \bibinfo {author} {\bibfnamefont {T.~A.}\ \bibnamefont
  {Brun}},\ }\href@noop {} {\bibfield  {journal} {\bibinfo  {journal} {arXiv
  preprint arXiv:1805.01599}\ } (\bibinfo {year} {2018})}\BibitemShut {NoStop}%
\bibitem [{\citenamefont {Bash}\ \emph {et~al.}(2015)\citenamefont {Bash},
  \citenamefont {Gheorghe}, \citenamefont {Patel}, \citenamefont {Habif},
  \citenamefont {Goeckel}, \citenamefont {Towsley},\ and\ \citenamefont
  {Guha}}]{qcovert1}%
  \BibitemOpen
  \bibfield  {author} {\bibinfo {author} {\bibfnamefont {B.~A.}\ \bibnamefont
  {Bash}}, \bibinfo {author} {\bibfnamefont {A.~H.}\ \bibnamefont {Gheorghe}},
  \bibinfo {author} {\bibfnamefont {M.}~\bibnamefont {Patel}}, \bibinfo
  {author} {\bibfnamefont {J.~L.}\ \bibnamefont {Habif}}, \bibinfo {author}
  {\bibfnamefont {D.}~\bibnamefont {Goeckel}}, \bibinfo {author} {\bibfnamefont
  {D.}~\bibnamefont {Towsley}}, \ and\ \bibinfo {author} {\bibfnamefont
  {S.}~\bibnamefont {Guha}},\ }\href@noop {} {\bibfield  {journal} {\bibinfo
  {journal} {Nature Communications}\ }\textbf {\bibinfo {volume} {6}} (\bibinfo
  {year} {2015})}\BibitemShut {NoStop}%
\bibitem [{\citenamefont {Sheikholeslami}\ \emph {et~al.}(2016)\citenamefont
  {Sheikholeslami}, \citenamefont {Bash}, \citenamefont {Towsley},
  \citenamefont {Goeckel},\ and\ \citenamefont {Guha}}]{qcovert2}%
  \BibitemOpen
  \bibfield  {author} {\bibinfo {author} {\bibfnamefont {A.}~\bibnamefont
  {Sheikholeslami}}, \bibinfo {author} {\bibfnamefont {B.~A.}\ \bibnamefont
  {Bash}}, \bibinfo {author} {\bibfnamefont {D.}~\bibnamefont {Towsley}},
  \bibinfo {author} {\bibfnamefont {D.}~\bibnamefont {Goeckel}}, \ and\
  \bibinfo {author} {\bibfnamefont {S.}~\bibnamefont {Guha}},\ }in\ \href@noop
  {} {\emph {\bibinfo {booktitle} {Information Theory (ISIT), 2016 IEEE
  International Symposium on}}}\ (\bibinfo {organization} {IEEE},\ \bibinfo
  {year} {2016})\ pp.\ \bibinfo {pages} {2064--2068}\BibitemShut {NoStop}%
\bibitem [{\citenamefont {Wang}(2016)}]{qcovert3}%
  \BibitemOpen
  \bibfield  {author} {\bibinfo {author} {\bibfnamefont {L.}~\bibnamefont
  {Wang}},\ }in\ \href@noop {} {\emph {\bibinfo {booktitle} {Information Theory
  Workshop (ITW), 2016 IEEE}}}\ (\bibinfo {organization} {IEEE},\ \bibinfo
  {year} {2016})\ pp.\ \bibinfo {pages} {364--368}\BibitemShut {NoStop}%
\bibitem [{\citenamefont {Bradler}\ \emph {et~al.}(2016)\citenamefont
  {Bradler}, \citenamefont {Kalajdzievski}, \citenamefont {Siopsis},\ and\
  \citenamefont {Weedbrook}}]{qcovert4}%
  \BibitemOpen
  \bibfield  {author} {\bibinfo {author} {\bibfnamefont {K.}~\bibnamefont
  {Bradler}}, \bibinfo {author} {\bibfnamefont {T.}~\bibnamefont
  {Kalajdzievski}}, \bibinfo {author} {\bibfnamefont {G.}~\bibnamefont
  {Siopsis}}, \ and\ \bibinfo {author} {\bibfnamefont {C.}~\bibnamefont
  {Weedbrook}},\ }\href@noop {} {\bibfield  {journal} {\bibinfo  {journal}
  {arXiv preprint arXiv:1607.05916}\ } (\bibinfo {year} {2016})}\BibitemShut
  {NoStop}%
\bibitem [{\citenamefont {Arrazola}\ and\ \citenamefont
  {Scarani}(2016)}]{qcovert5}%
  \BibitemOpen
  \bibfield  {author} {\bibinfo {author} {\bibfnamefont {J.~M.}\ \bibnamefont
  {Arrazola}}\ and\ \bibinfo {author} {\bibfnamefont {V.}~\bibnamefont
  {Scarani}},\ }\href@noop {} {\bibfield  {journal} {\bibinfo  {journal}
  {Physical Review Letters}\ }\textbf {\bibinfo {volume} {117}},\ \bibinfo
  {pages} {250503} (\bibinfo {year} {2016})}\BibitemShut {NoStop}%
\bibitem [{\citenamefont {Klesse}(2007)}]{klesse2007approximate}%
  \BibitemOpen
  \bibfield  {author} {\bibinfo {author} {\bibfnamefont {R.}~\bibnamefont
  {Klesse}},\ }\href@noop {} {\bibfield  {journal} {\bibinfo  {journal}
  {Physical Review A}\ }\textbf {\bibinfo {volume} {75}},\ \bibinfo {pages}
  {062315} (\bibinfo {year} {2007})}\BibitemShut {NoStop}%
\bibitem [{\citenamefont {Klesse}(2008)}]{klesse2008random}%
  \BibitemOpen
  \bibfield  {author} {\bibinfo {author} {\bibfnamefont {R.}~\bibnamefont
  {Klesse}},\ }\href@noop {} {\bibfield  {journal} {\bibinfo  {journal} {Open
  Systems \& Information Dynamics}\ }\textbf {\bibinfo {volume} {15}},\
  \bibinfo {pages} {21} (\bibinfo {year} {2008})}\BibitemShut {NoStop}%
\bibitem [{\citenamefont {Wilde}(2013)}]{wilde2013quantum}%
  \BibitemOpen
  \bibfield  {author} {\bibinfo {author} {\bibfnamefont {M.~M.}\ \bibnamefont
  {Wilde}},\ }\href@noop {} {\emph {\bibinfo {title} {Quantum information
  theory}}}\ (\bibinfo  {publisher} {Cambridge University Press},\ \bibinfo
  {year} {2013})\BibitemShut {NoStop}%
\bibitem [{\citenamefont {Cover}\ and\ \citenamefont
  {Thomas}(2012)}]{cover2012elements}%
  \BibitemOpen
  \bibfield  {author} {\bibinfo {author} {\bibfnamefont {T.~M.}\ \bibnamefont
  {Cover}}\ and\ \bibinfo {author} {\bibfnamefont {J.~A.}\ \bibnamefont
  {Thomas}},\ }\href@noop {} {\emph {\bibinfo {title} {Elements of information
  theory}}}\ (\bibinfo  {publisher} {John Wiley \& Sons},\ \bibinfo {year}
  {2012})\BibitemShut {NoStop}%
\bibitem [{\citenamefont {Nielsen}\ and\ \citenamefont
  {Chuang}(2010)}]{nielsen2010quantum}%
  \BibitemOpen
  \bibfield  {author} {\bibinfo {author} {\bibfnamefont {M.~A.}\ \bibnamefont
  {Nielsen}}\ and\ \bibinfo {author} {\bibfnamefont {I.~L.}\ \bibnamefont
  {Chuang}},\ }\href@noop {} {\emph {\bibinfo {title} {Quantum computation and
  quantum information}}}\ (\bibinfo  {publisher} {Cambridge university press},\
  \bibinfo {year} {2010})\BibitemShut {NoStop}%
\bibitem [{\citenamefont {Audenaert}(2007)}]{audenaert2007sharp}%
  \BibitemOpen
  \bibfield  {author} {\bibinfo {author} {\bibfnamefont {K.~M.}\ \bibnamefont
  {Audenaert}},\ }\href@noop {} {\bibfield  {journal} {\bibinfo  {journal}
  {Journal of Physics A: Mathematical and Theoretical}\ }\textbf {\bibinfo
  {volume} {40}},\ \bibinfo {pages} {8127} (\bibinfo {year}
  {2007})}\BibitemShut {NoStop}%
\bibitem [{\citenamefont {Alicki}\ and\ \citenamefont
  {Fannes}(2004)}]{alicki2004continuity}%
  \BibitemOpen
  \bibfield  {author} {\bibinfo {author} {\bibfnamefont {R.}~\bibnamefont
  {Alicki}}\ and\ \bibinfo {author} {\bibfnamefont {M.}~\bibnamefont
  {Fannes}},\ }\href@noop {} {\bibfield  {journal} {\bibinfo  {journal}
  {Journal of Physics A: Mathematical and General}\ }\textbf {\bibinfo {volume}
  {37}},\ \bibinfo {pages} {L55} (\bibinfo {year} {2004})}\BibitemShut
  {NoStop}%
\end{thebibliography}%

\end{document}